\documentclass[]{aa}
\usepackage{graphicx,amssymb,natbib}
\usepackage{epsf,psfig,epsfig}
\usepackage[dvips]{color}
\usepackage{graphicx}
\usepackage{natbib}
\usepackage{graphicx}
\usepackage[latin1]{inputenc}
\usepackage[T1]{fontenc}

\def\x1720{XTE~J1720$-$318 }
\def\nh{N$_{\mathrm{H}}$ }

\def\eg{{\it e.g.}}
\def\etal{et~al.~}
\def\ie{{\em i.e. }}
\def\x1720{XTE~J1720-318}
\def\nh{N$_{\mathrm{H}}$}
\def\kir{$\chi^2_{red}$}

\begin{document}
\title{High-energy observations of the state transition of the X-ray 
nova and black hole candidate XTE~J1720-318}

\author{M. Cadolle Bel\inst{1}, J. Rodriguez\inst{1,2,3}, P. Sizun\inst{1},
 R. Farinelli\inst{4}, M. Del Santo\inst{5}, A.~Goldwurm\inst{1,6},
P. Goldoni\inst{1,6}, S. Corbel\inst{1,7}, A. N. Parmar\inst{8}, E.
Kuulkers\inst{8}, P.~Ubertini\inst{5}, F. Capitanio\inst{5}, J.-P.
Roques\inst{9}, F. Frontera\inst{4,10}, L.~Amati\inst{10}, N.~J.~Westergaard\inst{11}}

\offprints{M. Cadolle Bel : mcadolle@cea.fr}

\institute{Service d'Astrophysique, DAPNIA/DSM/CEA - Saclay, 91191
Gif-sur-Yvette Cedex, France \and Integral Science Data Center,
Chemin d'Ecogia, 16, CH-1290 Versoix, Switzerland \and CNRS FRE
2591, France \and Physics Department, University of Ferrara,
44-100, Ferrara, Italy \and IASF-CNR, Via del Fosso del Cavaliere
100, 00133 Roma, Italy \and Fédération de Recherche APC, 11 place
M. Berthelot, 75231, France \and Université Paris VII, France 
\and Research and Scientific Support
Department, ESA, ESTEC, Keperlaan 1, NL-2200 AG Noordwijk, The
Netherlands \and Centre d'Etude Spatiale des Rayonnements, CNRS,
Toulouse Cedex 4, France \and IASF-CNR Section of Bologna, Via
P.Gobetti 101, 40129 Bologna, Italy \and Danish Space Research
Institute, Juliane Maries Vej 30, Copenhagen 0, DK-2100, Denmark}

\date{Received ; accepted}
\authorrunning{M. Cadolle Bel et al.}
\titlerunning{High Energy Observations of XTE~J1720-318}

\abstract{We report the results of extensive high-energy
observations of the X-ray transient and black hole candidate
\x1720 performed with INTEGRAL, XMM-Newton and RXTE. The source,
which underwent an X-ray outburst in 2003 January, was observed in
February in a spectral state dominated by a soft component with a
weak high-energy tail. The XMM-Newton data provided a high column
density \nh~of $1.2\times 10^{22}$~cm$^{-2}$ which suggests that
the source lies at the Galactic Centre distance. The simultaneous
RXTE and INTEGRAL Target of Opportunity observations allowed us to
measure the weak and steep tail, typical of a black-hole binary in
the so-called High/Soft State. We then followed the evolution of the source outburst over several
months using the INTEGRAL Galactic Centre survey observations. The
source became active again at the end of March: it showed a clear
transition towards a much harder state, and then decayed to a
quiescent state after April. In the hard state, the source was
detected up to 200~keV with a power law index of
$\sim$~1.9 and a peak luminosity of $\sim$~7~$\times $~$10^{36}$~erg~s$^{-1}$
in the 20-200 keV band, for an assumed
distance of 8 kpc. We conclude that \x1720 is indeed a new member
of the black hole X-ray novae class which populate our galactic
bulge and we discuss its properties in the frame of the spectral
models used for transient black hole binaries. } \maketitle

\section{Introduction}
\indent X-ray Novae (XN), also called soft X-ray transients, are
low mass X-ray binaries where a compact object accretes at a very
low rate from a late type companion star (Tanaka~\&~Shibazaki
1996). Although they are usually in a quiescent state (and
therefore nearly undetectable), they undergo bright X-ray
outbursts, with typical recurrence periods of many years,
which last several weeks or even months before the source returns
to quiescence. Most of the XN are associated to dynamically proven
Black Holes (BH) and indeed the great majority of the known 18 Black
Hole Binaries (BHB) as well as of the 22 binary Black Hole
Candidates (BHC) are transients (McClintock~\&~Remillard 2003).
Because of large changes in the effective accretion rates that
occur during the XN outbursts and the very hard spectra they
usually display, these sources provide powerful probes of the
accretion phenomena and radiation processes at work in BH, and are
primary targets for high-energy instruments. Indeed, during their
outbursts, these sources often undergo changes in their spectral
and temporal characteristics; they may pass through the different
{\it spectral states} observed in BHB (Tanaka \& Lewin, 1975,
McClintock~\&~Remillard 2003). The two principal states of BHB are
the Low/Hard State (LHS) and the High/Soft State (HSS). In the
latter, the emission is dominated by a very soft (kT~$\sim$1~keV)
component generally interpreted as the thermal radiation from an
optically thick and geometrically thin accretion disc (Shakura \&
Sunyaev 1973). A weak and steep power law may also be present and
little variability is observed. In the LHS, the spectrum is rather
described by a hard power law with photon index in the range
1.5--2.0 and a break around 100 keV. The LHS is also characterized
by large timing variability. Intermediate states (namely the
intermediate state and the very high soft state) are also
observed, where both the soft and hard components are present; the
source displays a complicated pattern of timing properties. In
spite of the recent advances in the characterization of BHB
spectral states and modelling of the emission components, the
basic mechanisms which generate the state transitions and in
particular the origin of the hard component are not yet
understood. Detection and broad band studies of new BH systems is
therefore essential to acquire better statistics on the
phenomenology of the BHB spectral states and on the relations
between their emission components.\\
\indent Since XN probably follow the galactic stellar
distribution, they are concentrated in the direction of the bulge
of our Galaxy (with a higher density towards the centre).
The SIGMA gamma-ray telescope on board the GRANAT satellite, and
later the hard X-ray instruments on board Rossi XTE and Beppo SAX
discovered and studied several (about 10) BHC XN in the bulge.
INTEGRAL, the INTErnational Gamma-Ray Astronomy Laboratory
(Winkler \etal 2003) is a European Space Agency observatory
launched on 2002 October 17, carrying four instruments: two main
gamma-ray instruments, IBIS (Ubertini \etal 2003) and SPI
(Vedrenne \etal 2003), and two monitors, JEM-X (Lund \etal 2003)
and OMC (Mas-Hesse \etal 2003). The IBIS coded mask instrument is
characterised by a wide Field of View (FOV) of 29°$\times$~29°
(9°$\times$~9° fully coded), a point spread function of 12' FWHM
and it covers the energy range between 20 keV and 8 MeV. The SPI
telescope works in the range from 20~keV to 8~MeV with a FOV of
31° diameter (16° fully coded), an angular resolution of 2.5°
(FWHM) and  a typical energy resolution of 2.5 keV at 1.3 MeV. The
JEM-X~monitor provides spectra and images with arcminute angular
resolution in the 3 to 35 keV band, with a FOV of about 10°
diameter. Thanks to its instruments performances and to the survey
program specifically dedicated to the Galactic Centre (GC) region,
INTEGRAL allows the detection and study of the hard X-ray emission
from BH XN at large distances and at weaker flux levels than
before.\\
\indent \x1720 was discovered on 2003 January 9 with the All Sky
Monitor (ASM) on board RXTE as a transient source undergoing an
X-ray nova like outburst (Remillard \etal 2003). The source
1.2--12~keV flux increased to the maximum value of $\sim$~430 mCrab
in 2~days (see Fig.~\ref{LCTOT}); then its flux started to decay
slowly. Follow up observations with the Proportional Counter Array
(PCA) on board RXTE showed the presence of a 0.6~keV thermal
component and a hard tail. The spectral parameters and the source
luminosity were typical of a BH (Markwardt 2003) in the so-called
HSS. Soon after, a radio counterpart was identified with the VLA
and ATCA radio telescopes (Rupen \etal 2003; O'Brien \etal 2003),
leading to the estimate of the most precise position
$\alpha_{J2000}=17^\mathrm{h}19^\mathrm{m}58^\mathrm{s}.985$,
$\delta_{J2000}=-31^\circ45^\prime01^{\prime\prime}.109$~$\pm~
0^{\prime\prime}$.25. The detection of its infrared counterpart
(Nagata \etal 2003) provided a measure of the extinction which is
compatible with a location of \x1720 at large distance, probably
several kpc.\\
\indent \x1720 was observed by XMM-Newton, RXTE and INTEGRAL in
2003 February during dedicated Target of Opportunity (ToO)
observations. It was then observed by INTEGRAL during the surveys
of the GC region performed in March and April and again from 2003
August to October. We report here the results based from these
observations, starting with the description of the available data
and of the analysis procedures employed (Sect. 2). We then report
the results in Sect. 3 before discussing them in Sect.~4.
\section{Observations and Data Reduction}
\indent \x1720 was observed by XMM-Newton on 2003 February 20,
during a public 18.5~ks ToO. Preliminary analysis of these data
provided an improved X-ray position of the source
(Gonzalez-Riestra \etal 2003), confirming the association with the
radio and IR source. One week after, we performed an INTEGRAL ToO
observation of \x1720 which started on 2003 February 28 with a
176~ks exposure. The latter was conducted in coordination with a
RXTE ToO observation which lasted about 2~ks. The source was
further observed during the INTEGRAL Core Program during a series
of exposures dedicated to the GC survey, from March 25 to April 19
for a total of 551~ks observing time. Another 275 ks exposure on
the source has been accumulated during ToOs on H~1743-322 (Parmar
\etal 2003) in 2003 April. The field containing \x1720 has also
been extensively monitored (about 700 ks exposure time) during the
second part of the 2003 INTEGRAL GC survey.\\
\indent The log of the observations and data used in this work is
summarized in Table~\ref{tab:log}. Figure~\ref{LCTOT} (top panel)
shows the 1.2--12 keV RXTE/ASM light curve of~\x1720
and also indicates the
intervals covered by the dedicated XMM-Newton, RXTE and INTEGRAL
observations discussed here.
\begin{table*}[htbp]
\begin{center}
\caption{\label{tab:log}Log of the \x1720 observations analysed in
this paper. }
\begin{tabular}[h]{lllll}
\hline \hline
Spacecraft & Observation Period & Exposure & Instruments & Observation type\\
& Dates (2003)~(\# revolution) & (ks) & & /Mode\\
\\
XMM-Newton & 02/20 & 18.5 & EPIC-PN & ToO/Small Window\\
Rossi-XTE & 02/28 & 2 ks & PCA& ToO\\
INTEGRAL & 02/28~-~03/02 (46) & 176 & JEM X-2~+~IBIS & ToO~$^a$\\
INTEGRAL & 03/15~-~04/03 (51~-~57) & 361 & IBIS & GCDE \\
INTEGRAL & 04/06~-~04/22 (58~-~63) & 175 & IBIS~+~SPI & ToO on H 1743-322~$^a$\\
INTEGRAL & 04/12~-~04/19 (60~-~62) & 191 & IBIS~+~SPI & GCDE \\
INTEGRAL & 08/02~-~10/16 (103~-~122) & 700 & IBIS & GCDE \\
\hline
\end{tabular}
\end{center}
Notes: ~a) 5$\times$5 dithering pattern around the target.
\end{table*}

\subsection{XMM-Newton Data Analysis}
We present here the data taken with the EPIC-PN camera on board
XMM-Newton. The PN camera was operating in Small Window mode. We
processed the data using the {\tt Scientific Analysis System}
v5.4.1 and the calibration files updated at the end of 2003 March.
We first filtered our data for background flares. Since \x1720 was
bright at the date of the observation (resulting in a strong pile
up in the PN camera), we adopted the selection criteria suggested
by Guainazzi (2001) to obtain the source spectrum. We extracted
the single events from an annulus with an internal radius of
15$^{\prime\prime}$, and an outer radius of 29$^{\prime\prime}$
around the position of \x1720. As we only used single events in
this corona, the effective exposure time of the extracted
spectrum was about 6~ks.\\
\indent We obtained the background spectrum from a sky region far
from the source and we built the response matrix (RMF) and
ancillary response (ARF) files consistent with the selections.
Adding 2.5$\%$ systematics, the resultant spectrum was then fitted
with
{\tt XSPEC} v11.3.0 (Arnaud 1996) between 0.7 and 11 keV.\\

\subsection{Rossi XTE Data Analysis}
We reduced and analysed the RXTE data with the {\tt LHEASOFT} package
v5.3. We reduced the data from the PCA following the standard
methods explained in the ABC of RXTE and the cook book. The good
time intervals (GTI) were defined when the satellite elevation
was~$>$~10$^\circ$ above the Earth's limb, and the offset
pointing~$<$~0.02$^\circ$. We also chose to retain the data taken
when most of the Proportional Counter Units (PCU) were turned on
(a maximum of 5 here). We extracted the spectra from the standard
2 data, from the top layer of each PCU. Background spectra were
produced with {\tt pcabackest}  v3.0, using the latest calibration files
available for bright sources. The RMF and ARF were generated with
{\tt pcarsp} v8.0. Due to uncertainties in the PCA RMF, we
included some systematic errors in the spectra. To estimate the
level of those systematics, we reduced and analysed a contemporary
Crab observation. To obtain a reduced $\chi^2$ of 1 when fitting
the Crab spectra, we set the level of systematics as follows:
0.6$\%$ between 2 and 8 keV and 0.4$\%$ above 8 keV. We fitted the
spectra between 3--25 keV for the PCA. We also processed HEXTE
data but, due to the poor statistics, we did not include the few HEXTE
data points in the analysis.\\
\indent  For the timing analysis, we extracted 16 s resolution PCA
light curves from standard 2 data, using all PCUs and all layers,
between 2 and 20 keV (absolute channels 5--49), and corrected them
for background. We extracted high temporal resolution light curves
from the event mode data with a nominal resolution of
$2^{-13}$~s~(\ie $\sim125$~$\mu$s) rebinned to 1~ms during the
extraction processes. Three such light curves were extracted
between absolute channels 5--49 (2--20 keV epoch 5), 17--49 (7--20
keV) and 24--49 (10--20 keV).\\

\subsection{INTEGRAL Data Analysis}
An INTEGRAL observation is made of several pointings (science
windows, hereafter SCW) each having exposure time lasting from
1800 to 3600~s and following a special pattern on the plane of the
sky (Courvoisier \etal 2003). Except for the $5 \times 5$
dithering mode for ToOs, the entire GC region was observed in the
framework of the Galactic Centre Deep Exposure (GCDE) program
(Winkler 2001). Deep exposures in the GC radian ($\pm$~30~deg in
longitude,~$\pm$~20~deg in latitude centred at \emph{l}=0,
\emph{b}=0) are obtained with a set of individual pointings
lasting
30 min each on a regular pointing grid.\\
\indent All the INTEGRAL instruments were operating
simultaneously. We describe here mainly results obtained from the
data recorded with the ISGRI detector (Lebrun \etal 2003) of the
IBIS telescope covering the spectral range from 20 to 800 keV. For
the first observation set, when the source was very soft, we also
present data from the JEM-X instrument. The IBIS data have been
reduced with the {\tt Offline Scientific Analysis (OSA)} v3.0
delivered in December 2003 to produce images and extract spectra
for each SCW (Goldwurm \etal 2003). We selected SCW for which the
source was within 8$^\circ$ from the telescope axis. For the
spectral analysis, we used a 12 linearly rebinned channel RMF and
the associated recently corrected ARF
(P. Laurent, 2003 December, private communication).
The resultant spectrum was fitted between 20 and 600 keV, but
above 200 keV the source is not always significantly detected and
below 20 keV systematic uncertainties are still very high.
Systematics errors at level of 8$\%$ (see Sect. 3.3) were applied
in the spectral fits to account for the residual effects of the
response matrix (Goldwurm \etal 2003). For the image analysis, the
background derived from empty fields was subtracted before
deconvolution and we used a catalog of about 41 sources to analyse
the images. The total amount of IBIS data we processed was
equivalent to about 1700 ks of exposure time, however due to
selections performed and the fact that the source was very often
off-axis, the effective exposure time is reduced to 652 ks.\\
\indent We reduced the JEM-X data with {\tt OSA} v3.0. Only the
JEM-X2 monitor was operated during our observation. Because of
uncertainties in the RMF for high off-axis angles, we selected
only SCW where the source was closest to the centre of the field
of view (\ie $<~$3$^\circ$ from the telescope axis). The energy
channels were also rebinned so as to have a $\sigma$~$>$~3. When
all these conditions hold, we extracted the spectra for an
effective exposure time of 21~ks. We fitted the resultant averaged
spectrum between 3.5 and 26.5 keV,
with the standard RMF and ARF.\\
\indent Concerning SPI, the data obtained during revolutions (hereafter, rev.) 58 and 60 to 63 were reduced using the
\emph{Spi\_science\_analysis\_2} script (Kn\"{o}dlseder 2004)
available in {\tt OSA} v3.0 (February 2004). Images were then
extracted between 20 and 40~keV to build a catalogue of sources.
Spectra were finally extracted for each data set with 50
logarithmic bins in the 20-1000~keV energy range, using {\tt Spiros}
(Skinner \& Connell 2003).
We used the imaging results from the IBIS/ISGRI telescope to
determine the active sources of the region in order to account for
their contribution in the SPI spectral extraction. The latter was
performed with a background model derived from the evolution of
the saturated count rates in the Ge detectors while the ratios
between detectors were left free in the analysis. A few SCW with a
bad $\chi^2$ were excluded. Extracted with {\tt Image Response Files}
v15, the resulting spectra were fitted with the RMF delivered in
2004 February.

\section{Results of the Analysis}
\subsection{The INTEGRAL Detection of the \x1720 State Transition}
\begin{figure}
\begin{center}
\includegraphics[width=1.\linewidth,keepaspectratio]
{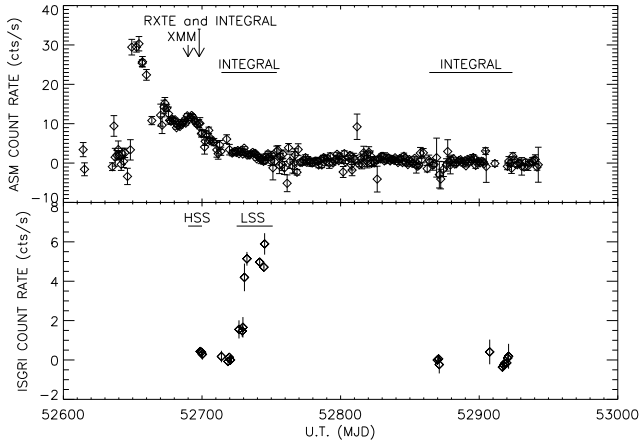}\end{center} \caption{\label{LCTOT} {\bf{Top: }}
The RXTE/ASM daily average 1.2-12~keV light curve of
\x1720 from few days before the outburst to 2003 October.
The arrows show the dates of the XMM-Newton,
RXTE and INTEGRAL observations. The approximate periods of later
INTEGRAL observations are indicated by horizontal lines. Universal
time is reported in units of MJD.
{\bf{Bottom: }}
The 20-120~keV IBIS/ISGRI light curve of \x1720 with
time bins of 2 days (rev. 46 to 122).
The data used to build spectra for HSS
and LHS are indicated by horizontal
lines.}
\end{figure}

\begin{figure}
\begin{center}
\includegraphics[width=1.\linewidth,keepaspectratio]{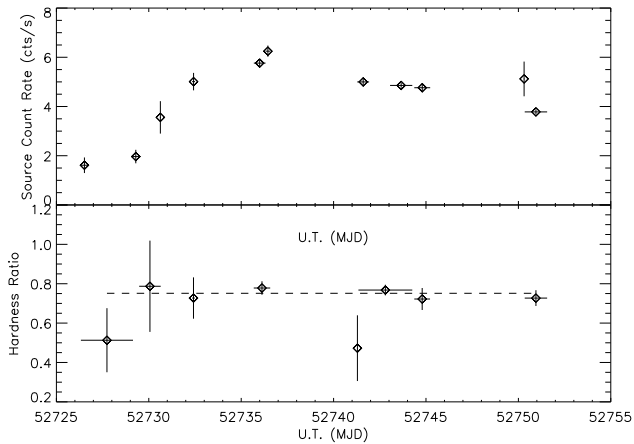}\end{center}
\caption{\label{HR}{\bf {Top: }} IBIS/ISGRI 20--80 keV light curve
with time bins $\sim$~1 day during the \x1720 hard flare (rev. 55 to 63). {\bf{Bottom: }} Corresponding hardness ratio, defined
as the ratio between the source count rate in the 40--80 keV band
and in the 20--40 keV band with time bins of 3 days. The dashed
line represents the average HR.}
\end{figure}

\begin{figure}[htbp]
\begin{center}
\includegraphics[width=0.7\linewidth,angle=270,keepaspectratio]{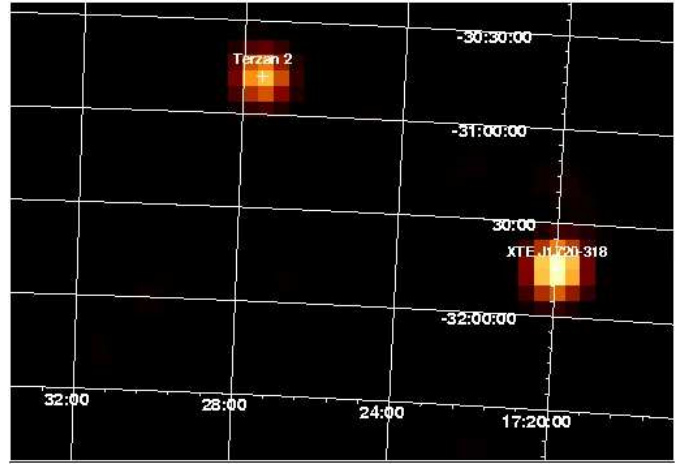}\end{center}
\caption{\label{Mosa}The IBIS/ISGRI reconstructed sky image of the
region around \x1720 in the 20--60 keV band
(rev. 58 to 61). \x1720 appears at a significance level of
94~$\sigma$ over the background. The other source in the image is
the hard X-ray burster located in the globular cluster Terzan~2.}
\end{figure}

The 20--120~keV IBIS/ISGRI light curve of \x1720 from the whole
set of the 2003 INTEGRAL data is shown in Fig.~\ref{LCTOT}
(lower panel) on the same scale of the ASM light curve. During
the INTEGRAL ToO observation of February 28 (rev. 46, MJD 52699), the source was detected at a very low flux level
above 20~keV and 10 days later (i.e. between March 9 and 20), it had decreased below the detection level. Starting from
March 25 (MJD 52724, rev. 54), the source
appeared to brighten in the INTEGRAL/IBIS energy band. Since a
similar behaviour was not seen in the ASM light curve
(see Fig.~\ref{LCTOT}, top panel), Goldoni
\etal (2003) proposed that the source was entering a hard state.
Figure~\ref{HR} shows the details of this hard flare.
The 20--80 keV flux was at the beginning
around 2~cts~s$^{-1}$ ($\sim$~11~mCrab) and increased to a maximum
level of 6.25~cts~s$^{-1}$ ($\sim$~34.5~mCrab) on April 6
(rev. 58, MJD 52737). After this, the flux gradually
decreased to the value of 4~cts~s$^{-1}$ (rev. 63, MJD
52751). When the INTEGRAL GC survey included the source again in
the IBIS FOV in mid August 2003, the transient was not detected
and remained below the IBIS/ISGRI detection level for the rest of
2003 (Fig.~\ref{LCTOT}). The derived 3~$\sigma$ upper limit on \x1720 flux
during the mid-August observations ($\sim$ MJD 52869, exposure of 271 ks) is
1.5 mCrab between 20--80 keV.\\
\indent Figure~\ref{HR} shows the hardness ratio (HR) measured
during the observed increase of the source high energy flux. 
There is no significant variation in the HR around its mean value of
0.75, only a slight indication of a softer HR ($\sim$~0.5) at the
beginning of the flare. We therefore used the whole data of this
hard flare to build up an average spectrum (see Sect. 3.3.2).
Moreover, we could add the data taken with SPI, which provide us
significant points during this period at high energies. We also
analysed JEM-X data taken during the hard flare, but due to the
fact that the source was often at large off-axis angles and very
faint below 20 keV, the derived data points were not significant; 
therefore, we did not use them in the analysis.\\
\indent Assuming an exponential shape for both the rise and the
decay phases of the hard flare we obtained, using the IBIS/ISGRI
data points of Fig.~\ref{LCTOT} (lower panel), time constants of 13 days 
(rise) and 48 days (decay) respectively. Since we used the August upper
limits for this estimate, the characteristic decay time we derived
is therefore only an upper limit. However, the hard flare
timescales appear comparable to the main outburst (the
characteristic decay time is 60 days) even if the peak broad-band
X-ray luminosity remains well below the peak luminosity of the
main outburst (see Sect. 3.3).
\\
\indent In the combined IBIS/ISGRI images obtained during the hard
outburst (data from rev. 58 to 61), \x1720 is detected at
94$\sigma$ in the 20-60 keV range (Fig.~\ref{Mosa}). The best
position found with IBIS from the 20-60 keV image is
$\alpha_{J2000}=17^\mathrm{h}19^\mathrm{m}58^\mathrm{s}$.7,
$\delta_{J2000}=-31^\circ44^\prime43^{\prime\prime}$.7 with an
accuracy of $0^\prime$.45 at $90\%$ confidence level (Gros \etal
2003). This position is consistent with the most precise position
of \x1720 derived from radio data since the offset is only
$17^{\prime\prime}$.7. The high-energy source is therefore
unambiguously associated to the transient.
\begin{figure}[htbp]
\epsfig{file=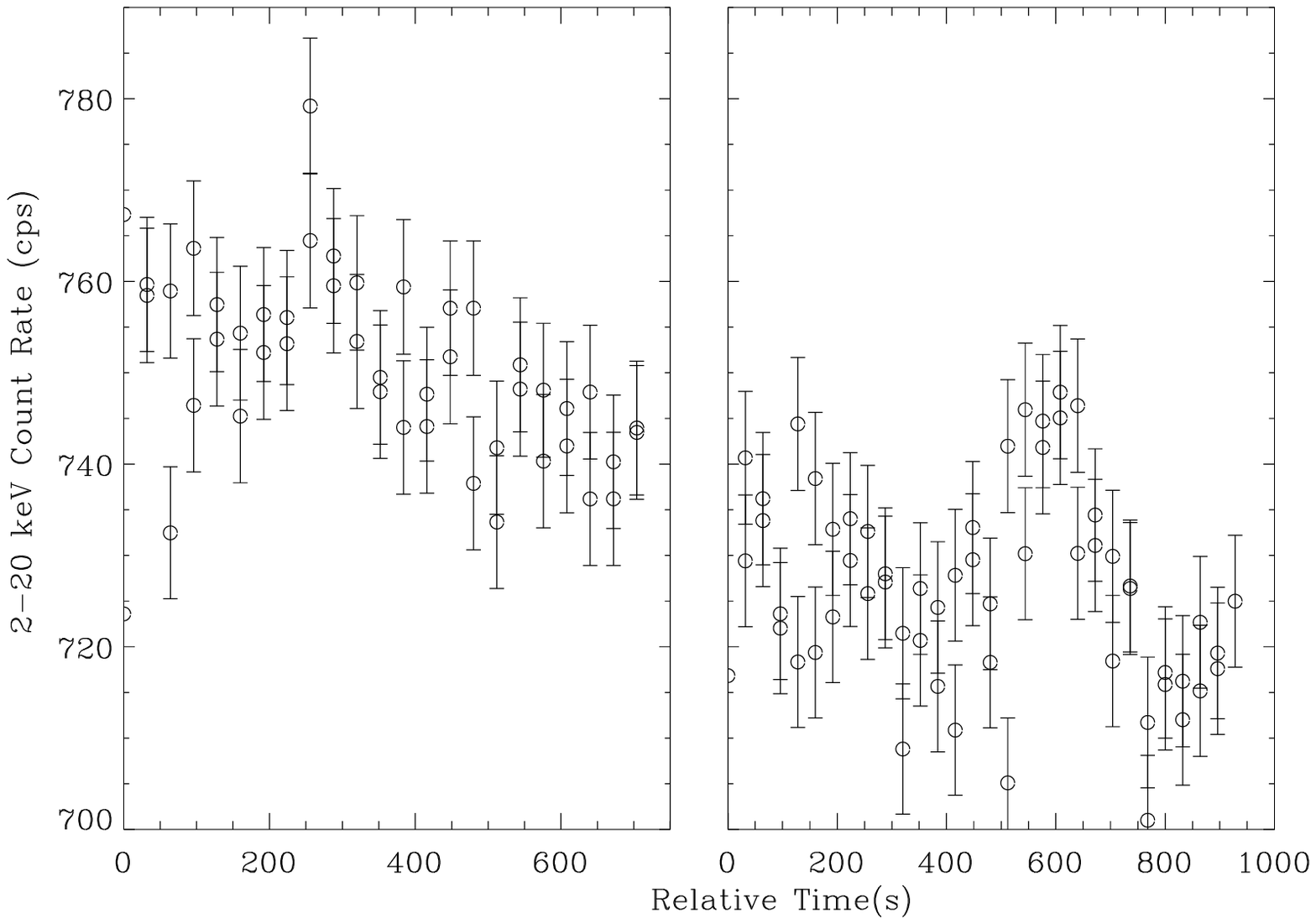, width=9.5cm}\\
\epsfig{file=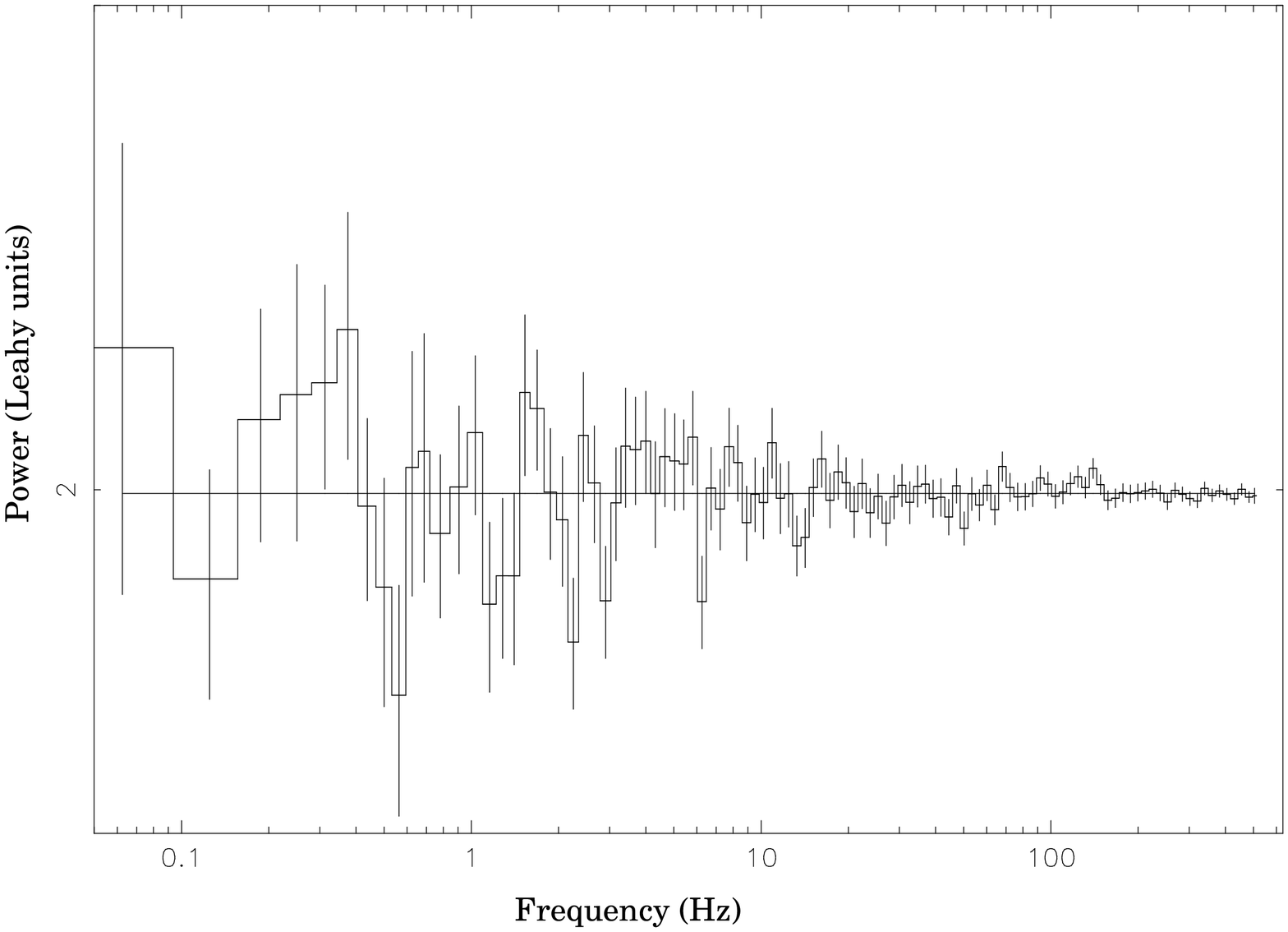,width=\columnwidth}\\
\caption{\label{pca} {\bf {Top: }} 2--20 keV RXTE/PCA light curves in counts
per second, covering the two data sets. For both, relative time 0
refers to the beginning of the GTIs: time 0 corresponds to
MJD=52698.506(28) (left panel) and to MJD=52699.487(31) (right
pannel). The time sampling is 16~s. {\bf{Bottom: }} 2--20 keV PDS
of the combined RXTE/PCA data sets. The best-fit (constant) is
superimposed as a line.}
\end{figure}
\subsection{\x1720 Timing Variability during the High/Soft State}
The XMM-Newton and the INTEGRAL/RXTE observations of February 2003
caught the source in a very soft state (HSS). The source appeared bright
at low energies, with a daily-averaged flux between 100 and
140~mCrab in the 1.2--12 keV band. The JEM-X and PCA instruments
detected the source at very high significance and we could derive
significant spectra up to 20 keV (Fig.~\ref{SIMULT}). However, the high-energy
emission was quite weak. IBIS detected the source at a level of
0.4~$\pm$~0.07~cts~s$^{-1}$ ($\sim$~2.1~mCrab) in the 20--120 keV
band with a signal of 6~$\sigma$, providing only few data points
at energies higher than 50 keV. As RXTE/HEXTE provided low
significant data points at energy $\geq$~20~keV, we did not
include them in the spectra (described in Sect.~3.3.1).\\
\indent The PCA 2--20~keV light curves are shown on Fig.~\ref{pca}.
\x1720 shows some variations around a mean value of $\sim
750$~cts~s$^{-1}$ in the first set, and $721$~cts~s$^{-1}$ in the
second set. A slight decrease is visible from the first
observation to the second (Fig.~\ref{pca}). The light curve of
the latter is characterised by an increase of the flux to $\sim
750$~cts~s$^{-1}$, during a $\sim 200$~s small flare. We produced
Power Density Spectra in the 3 energy ranges described in Sect.
2.2 with {\tt POWSPEC} v1.0. These energy dependant PDS were
produced on interval length of 16~s between 62.5 mHz and 500 Hz.
All the intervals (from the 2 data sets) were averaged in a single
frame, and a geometrical rebinning has been applied. The 2--20~keV
Leahy normalised PDS is flat (Fig.~\ref{pca}, lower panel).
The best-fit model is a constant value of 1.993~$\pm$~0.004 (at
the 90$\%$ confidence level) with a $\chi^2$ of 87.6 (105 degrees
of freedom, hereafter d.o.f.). This value is compatible with the
expected Leahy normalised value of 2 for purely Poisson noise
(white noise). At higher energy the PDSs are also flat. The
3~$\sigma$ upper limits on the 2--20 keV fractional level of
variability is $\sim6.7\%$. In the higher energy ranges, the upper
limit is rather high and meaningless due to the low statistics of
the source.

\subsection{Spectral Results}
\subsubsection{The High/Soft State Spectrum}
\begin{figure}
\epsfig{file=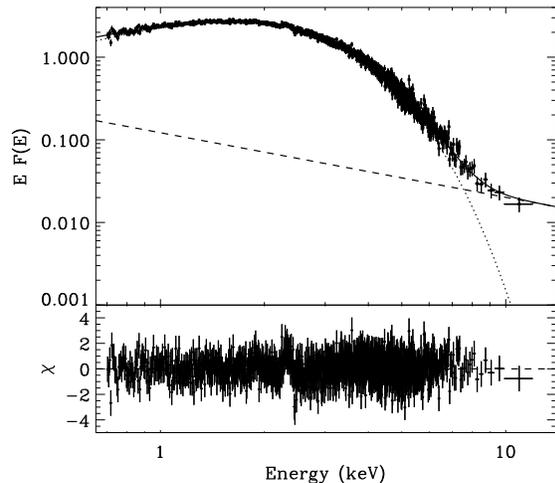, width=0.9\linewidth}\\
\caption{\label{XMM} XMM-Newton/EPIC-PN unabsorbed EF(E) spectrum of \x1720 (units keV~cm$^{-2}$~s$^{-1}$)
along with the best-fit model MCD plus power law.~{\it Dotted}:
MCD.~{\it Dashed}: power law.~{\it Thick}: total model.
Residuals (in $\sigma$ units) are also shown.}
\end{figure}

\begin{figure}[htbp]
\begin{center}
\includegraphics[width=0.65\linewidth,angle=270,keepaspectratio]{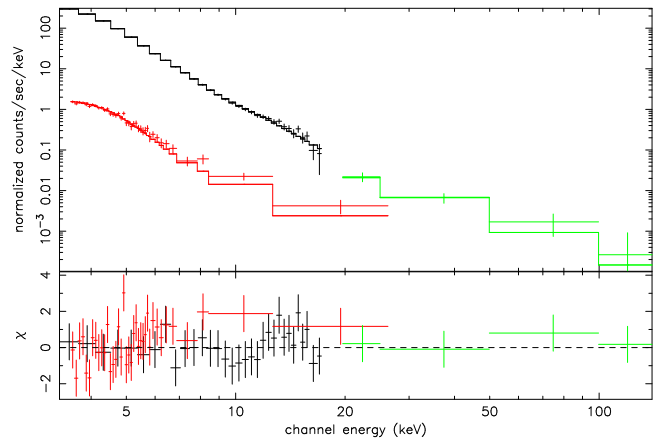}\end{center}
\caption{\label{SIMULT} Joint RXTE/PCA (black), INTEGRAL/JEMX-2
(red) and INTEGRAL/IBIS (green) spectra of \x1720 during the
observations of end 2003 February. The best-fit model, an absorbed
MCD plus a power law, is over plotted as a solid line to the data.
Residuals (in $\sigma$ units) are also shown.}
\end{figure}

\begin{figure}
\begin{center}
\includegraphics[width=0.75\linewidth,angle=270]{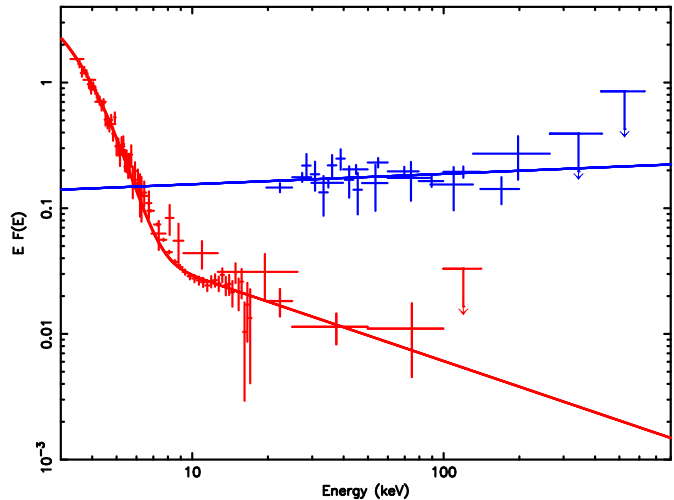}\end{center}
\caption{\label{pho} Unabsorbed EF(E) spectrum of \x1720
(units of keV~cm$^{-2}$~s$^{-1}$)
along with the best-fit model: MCD plus power law
for the HSS (red) and a power law for the LHS (blue) with IBIS/ISGRI and SPI 20--600 keV data (rev. 55 to 63).
Upper limits are shown at 3$\sigma$.}
\end{figure}

We have fitted the XMM-Newton EPIC-PN data with a model
composed of an absorbed multi-colour black-body disc (MCD, Mitsuda
\etal 1984) plus a power law. A single absorbed MCD alone leads to
a poor fit ($\chi^{2}$~=~1349 for 1064 d.o.f.), as does a single
absorbed power law ($\chi^{2}$~=~14825 for 1064 d.o.f.). The
best-fit parameters derived from our analysis are given in
Table~\ref{tab:XMM-RXTE-INT}. We obtained for
\nh~=~(1.24~$\pm$~$0.02$)~$\times$~$10^{22}$~cm$^{-2}$. The
unabsorbed flux in the 0.7--10 keV range is
$6.43\times10^{-9}$~erg~cm$^{-2}$~s$^{-1}$. Assuming a distance of
8 kpc (see discussion), we derive a 0.7--10 keV unabsorbed)
luminosity of $4.9$~$\times$~$10^{37}$~erg~s$^{-1}$. The disc component
accounts for more than $85\%$ of the total 2--100 keV luminosity.
If we assume a line of sight inclination angle ($\theta$) of 60$^\circ$,
we find, from the disc normalisation, an inner disc radius
of 48.7~$\pm~0.5$~km. Figure~\ref{XMM}
shows the resultant EF(E) spectrum and its best-fit.\\
\begin{table*}[htbp]
\begin{center}
\caption{\label{tab:XMM-RXTE-INT} \x1720 best-fit spectral
parameters (with 90~$\%$ confidence
level errors) for the XMM-Newton and the
RXTE/INTEGRAL ToOs of February.}
\begin{tabular}[h]{lllllll}
\hline \hline
Satellite&Date&Photon&Disc Tempe-&Disc Norma-&\kir&Flux~$^b$\\
&(2003)&Index&rature~(keV)&lisation~$^a$&(d.o.f.)&($\times$10$^{-9}$erg~cm$^{-2}$~s$^{-1}$)\\
\\
XMM-Newton & 02/20&$2.81~_{-0.66}^{+0.60}$&0.67~$\pm~0.01$&$1855~_{-43}^{+37}$&$0.84~(1062)$&$2.36$\\
RXTE+INTEGRAL & 02/28-03/02&$2.72~_{-0.34}^{+0.29}$&0.59~$\pm~0.01$&$5647~_{-404}^{+338}$&$0.97~(72)$&$3.25$\\
\hline
\end{tabular}
\end{center}
Notes:
a) Disc normalisation K is given by K~$=~(\frac{R}{D})^{2}\times\cos{\theta}$
where R is the inner disc radius in units of km, D is
the distance to the source in units of 10 kpc and $\theta$ the
inclination angle of the disc.\\
b)Unabsorbed 2--100 keV flux.
\end{table*}
We have applied the same absorbed MCD plus a power law model to a
simultaneous fit of the RXTE/PCA, INTEGRAL/JEM-X and INTEGRAL/IBIS
data taken about 8 days later. We obtained the best-fit parameters
reported in Table~\ref{tab:XMM-RXTE-INT}. To account for
uncertainties in relative instruments calibrations, we let a
multiplicative constant free to vary in the fit of the different
data sets. Taking the RXTE/PCA spectrum as the reference, the
derived constants are all found very close to 1 for each
instrument. As RXTE and JEM-X are not suited to determine
interstellar absorption (energy lower boundary is $\sim$~3~keV),
we fixed \nh~to the value obtained from the XMM-Newton fits.
We
also added a gaussian function at the iron fluorescent line energies
to account for a feature present in the RXTE data. The line
centroid was found to be $6.45~_{-0.35}^{+0.16}$ keV with an
equivalent width (EW) of $572~_{-178}^{+307}$~eV. However, this
line was not present in the data obtained with XMM-Newton. To check the reality of this line, we reperformed the fit of the EPIC PN spectrum by adding to the best fit continuum model an iron line at a fixed energy and width equal to the ones found from the RXTE data (FWHM = 1.6 keV). We obtained an upper limit for such a line of 250~eV EW at the 90$\%$ confidence level. For a narrow line at the same energy, we obtained an upper limit of 75~eV EW. This upper limit suggests that the line seen with RXTE is probably due to an incorrect background subtraction and not to \x1720. Indeed a contamination by the galactic ridge emission (Revnivtsev 2003) cannot be excluded even if the line should be rather centered at 6.7 with a narrower width. With a line centroid fixed to this energy, we obtained from the RXTE
spectrum a line width of $0.59~_{-0.21}^{+0.06}$~keV and an
equivalent width (EW) of $456~_{-136}^{+117}$~eV. The residuals are
slightly worse around the 6.4-6.7 range but they do not
exclude such a line contamination.
Detailed analysis of other RXTE/PCA data of \x1720 during the outburst
will probably clarify this issue. For this reason, we did not included the line for the fit of the
INTEGRAL data.
In spite of the low significance level of the
detection, the IBIS/ISGRI data allow us to study the source up to
higher energies because of the higher sensitivity of ISGRI and the
longer exposure time.
The 3-200 keV count spectrum and the derived
best-fit model are shown in Fig.~\ref{SIMULT}. Figure~\ref{pho}
(red) shows the unfolded EF(E) spectrum with its best-fit model.
Note that above $\sim$~100~keV, the source is not significantly detected.\\
\indent The disc inner radius (with the same assumptions on
distance and viewing angle as above) is 85~$_{-4}^{+2}$~km and
the disc flux luminosity contributes to 93$\%$ of the unabsorbed
2-100~keV luminosity. Indeed, there is a slight evolution between the
XMM-Newton derived disc parameters (radius and temperature) and the
same parameters found one week later by RXTE and INTEGRAL
while the power law slope did not change.
According to the strengths of the soft component and the value of
the power law photon index, we found that the source was clearly
in a HSS, where the thermal component from the accretion disc
dominates and the high energy tail is very weak. We have also
fitted the data with a bulk motion comptonisation model (Shrader
\& Titarchuk 1999), often used to model the spectra of BH in HSS.
The test gave us an acceptable fit with $\chi^2_{\nu}$ of 1.36 for
78 d.o.f..
The derived temperature of thermal photon source is
$0.52~\pm~0.01$~keV, the energy spectral index $1.8~_{-0.4}^{+0.2}$
and the log~A parameter $-1.6~_{-0.2}^{+0.1}$, compatible with the parameters obtained from
the MCD plus power law fit.
The spectra taken
during the last week of 2003 February are therefore all consistent
with the hypothesis that \x1720 is a BH XN in HSS.

\subsubsection{The Low/Hard State Spectrum}
As discussed above, IBIS data from rev. 55 to 63, during the
hard flare, are consistent with one another (\ie no variation of
HR) and can be summed to derive the average spectrum and its
best-fit model reported in Fig.~\ref{pho} (blue).
We fitted this spectrum
with a simple power law model between 20 and 600 keV.
We also used the count spectrum derived from SPI data of
rev. 58 to 61 in order to make a simultaneous fit.
Due to the presence of high background structures,
SPI data from rev. 62 to 63 were not included in the spectra.
The SPI data points were binned
so as to have a level of 3~$\sigma$ per bin or 4 bins together at
least. To account for uncertainties in relative instruments
calibrations, we let a multiplicative constant vary in the fit.
Taking the IBIS spectrum as the reference (constant equal to 1),
the multiplication factor returned for SPI during the fit is 1.28.
At 90$\%$ confidence level, the best-fit photon index returned from
the fits is 1.9$~\pm~0.1$ with a reduced $\chi^2$ of 1.60 (22~d.o.f.),
which reveals that the spectrum of \x1720 is much
harder than observed in February. In addition to the power law
model, we fitted the data set with a comptonisation model (Sunyaev
\& Titarchuck 1980): the XSPEC {\it compst} model.
The derived parameters~are $43~_{-11}^{+32}$~keV for the temperature
and $2.7$~$\pm~0.9$ for the optical depth,
with a reduced $\chi^2$ of 1.27 (21~d.o.f.).
Similarly we fitted the spectrum with a comptonisation model
(the XSPEC {\it comptt} model)
which includes relativistic effects, estimates a larger range of
parameters and includes the seed soft photon energy (Titarchuk 1994).
We obtained, with a seed photon temperature frozen to kT$_{bb}$~=~0.6 keV
and a spherical geometry, a plasma temperature of kT~=~$57~\pm~29$~keV
and an optical depth $\tau$~=~2~$\pm~1$ with a reduced $\chi^2$ of 1.19
(21~d.o.f.). We also tested the Putanen \& Svensson (1996) comptonisation model (the XSPEC {\it compps} model) proper for very hot plasmas: we obtained kT~=~$430~\pm~110$~keV and $\tau$ equal to $0.11~\pm~0.11$, for a fixed seed photon kT$_{bb}$~=~0.45,
with a reduced $\chi^2$ of 1.33. Since the latter models gave slightly better fit than the single power law,
we have performed an additional test to see if a
break in the power law would be statistically significant.
We fitted a cut-off power law and derived the difference between the
absolute $\chi^2$ with the $\chi^2$ of the single power law. We restricted the fit to the data up to 300 keV to avoid the use of upper limits,
and we obtained a $\Delta\chi^2$ of 5.8; for a $\chi^2$ distribution
with 1 d.o.f., this value represents a probability of 95$\%$
that the new component is significant.
Even if the test is not fully conclusive, a cut-off in the model with a typical folding energy of approximately 120 keV clearly improves
the fit and better describes the available data.
We note that the derived thermal comptonisation
parameters for \x1720 during the hard flare (or the best-fit power
law index and cut-off energy) are very much consistent with those
found in BHB in the so-called LHS.

\section{Discussion}
The high equivalent absorption column density derived from the
XMM-Newton data suggests that \x1720 lies at the GC distance or
even further. This would place the source in the galactic bulge
and we will, therefore, assume a distance to the source of 8~kpc.\\
\indent When observed with XMM-Newton, about 40 days after the
outburst peak, \x1720 was clearly in a HSS, characterized by a
strong soft (thermal) component, well modelled by a MCD model with
an inner disc temperature of kT~$\sim$~0.7~keV, and a weak power
law tail (Fig.~\ref{XMM}). The source was found in HSS also at the end of February
(Fig.~\ref{SIMULT} and Fig.~\ref{pho}) when we could measure, with higher precision, using
INTEGRAL and RXTE simultaneous observations, the power law index of 2.7.
In both observations, the disc component
accounted for more than 85$\%$ of the unabsorbed 2--100 keV source
luminosity, estimated at the end of February at
2.5~$\times~10^{37}$~erg~s$^{-1}$. We estimated the bolometric luminosity from the best fit spectrum by extending the flux computation at 0.01 keV. We obtained for the XMM observation a value of 1.4~$\times~$10$^{38}$~erg~s$^{-1}$. Even for a small
5~M$_{\odot}$ BH (see discussion below), this bolometric luminosity
is below the Eddington luminosity which is 6.5~$\times$~10$^{38}$~erg~s$^{-1}$
for such a BH mass. Similar results are obtained one week later
with RXTE/INTEGRAL observations: the accretion rate is
sub-Eddington.\\
\indent Besides, no line emission was observed
with XMM-Newton and we could determine an upper limit to the EW of
 75 and 144~eV for narrow lines at 6.4 and 6.7 keV respectively.
The upper limits for broad lines are less constraining.
As discussed in Sect.~3.3.1, we consider unlikely that the
relatively strong iron line ($\sim$~570 eV) we detected with RXTE
about 10 days later can be due to the source since we
obtained an upper limit of 250 eV with XMM data
and we did not observe large
spectral changes between the two observations.
The XMM-Newton narrow line upper limit is below the strong
EWs ($>$ 150 eV) of lines observed in certain BH systems
and attributed to fluorescence produced by reflection
of hard X-rays from the accretion disc (\eg~Miller \etal 2001).\\
\indent However, these line are often broadened by relativistic effects
and in this case our data are less constraining.
For the parameters of the emission line
(centroid at 6.2 keV and FWHM of 2.4 keV)
reported by Markwardt (2003) from an RXTE observation on of \x1720
performed during the main outburst peak,
we in fact obtained upper limits of only 290~eV, while the
RXTE measured equivalent width was 95 eV.
Only slight differences in the spectral parameters of the
soft component were measured between the INTEGRAL/RXTE
observations and the XMM-Newton one (\ie a lower temperature and a
larger inner disc radius). The inner radius derived from
the normalisation of the MCD model for \x1720 (for the assumed
distance and inclination angle) is compatible with values of the
radius of the innermost stable circular orbit around a
Schwarzschild BH with mass of $\sim$ 5 M$_{\odot}$. During the
outburst decay of BH XN, this parameter is observed to remain
approximately constant if the source stays in the HSS (McClintock
\& Remillard 2003; Ebisawa \etal 1991). The variation observed in
\x1720 could indicate that the disc was receding during the decay
phase, but it is more probably linked to a specific variation of
the normalisation during a secondary flare. Indeed the XMM-Newton
observation took place right at the maximum of a weak secondary
peak which occurred in the decay phase (see Fig.~\ref{LCTOT},
top panel) and which was also observed in infrared (Nagata \etal 2003). In anycase, these differences could also be due at least in part to
cross calibration uncertainties between the instruments.
The lack of significant rapid variability that we have found from
the study of the PDS using RXTE data, is also compatible with the
source being in the HSS. In this state in fact, the fractional
time variability is in general lower than 5-10$\%$. From our data,
considering the results previously reported by Markwardt \etal 2003 and
the RXTE/ASM light curve which steadily decreased during 2003, we
can conclude that the source did not change spectral state during
the decay phase which started after the main outburst peak and
lasted until about mid March.
\\
\indent A dramatic change in the source behaviour was instead
observed with INTEGRAL towards the end of March. After the source
had decreased below the INTEGRAL detection level, we observed the
rise of the high-energy component about 75 days after the main
outburst peak, giving rise to a secondary outburst which we could
observe for about 25 days. Since such increase was not seen in the
RXTE/ASM count rate, and as we did not have significant signal in
the JEM-X data, we conclude that the source underwent a spectral
transition towards the LHS. The 20--200~keV unabsorbed luminosity
increased in about 10 days from below the INTEGRAL detection level to a
value of 7~$\times10^{36}$~erg~s$^{-1}$.
Then it started to decrease
with timescale between 10 days and 50 days.\\
\indent During this secondary outburst, the $>$ 20 keV spectrum
was hard and well described by a power law photon index of 1.9 or
a thermal comptonisation model with a (weakly constrained) plasma
temperature of 43 keV and an optical depth of 2.7.
The spectral break seems probable, but no firm conclusion is possible
due to the low significance of the derived source spectrum at high
energies. The derived best-fit parameters (both the power law
slope and break and the temperature and depth of the comptonising
plasma) are however compatible with those typically found in BHB
in the LHS. Assuming that the power law extends at low energies
without any additional contribution of a soft component, we
estimated an average unabsorbed 2--200 keV luminosity of
$\sim$~9.4~$\times$~10$^{36}$~erg~s$^{-1}$. The bolometric luminosity is estimated to 4.3~$\times$~10$^{37}$~erg~s$^{-1}$
and shows again that the source was in a sub-Eddington regime, even
for a low mass BH.
While this secondary outburst did not reach the luminosity
of the main one, it is clear
that the transition is not simply due to spectral pivoting as
observed in Cygnus X-1.\\
\indent The high peak luminosity, the fast rise and slow decay
time scales, the HSS and the secondary outburst with transition to
a LHS with spectral parameters typically observed in other
(dynamically confirmed) BH transients, like \eg~XTE~J1550-564
(Sobczak \etal 2000; Rodriguez \etal 2003) or GRO~J1655-40
(Sobczak \etal 1999, see also McClintock \& Remillard 2003),
clearly show that \x1720 is very likely a new XN and BHC, possibly
located in the galactic bulge.
\\
\indent Although there is little doubt about the origin of the
soft thermal component and its modelling, the interpretation of
the high-energy tail and its connection to the spectral states
remain rather controversial. In the HSS, most of the X-rays are
radiated by the accretion disc which is supposed to extend down
very close to the BH horizon. The standard Shakura \& Sunyaev
(1973) $\alpha$-disc, however, cannot produce hard radiation (in
either of the spectral states). In the LHS, the disc component is
weak or absent and, when observed, the fitted temperature is very
low and the inner radius very large. In this state, the accretion
disc is supposed to be truncated at a large radius. The hard
component is generally attributed to thermal comptonisation of the
disc soft radiation by a hot corona (Sunyaev \& Titarchuk 1980;
Titarchuk 1994) located above the disc or in the inner part of the
system, around and very close to the BH.  Thermal comptonisation
models fit well the spectra during the LHS and indeed the
comptonisation parameters derived for \x1720 are in good agreement
with those usually observed in BH systems in LHS. However, the
details of the geometry and of radiation mechanisms at work are
still not understood; the processes which lead to the spectral
transition and the possible role of non-thermal (synchrotron)
radiation are still very uncertain. For example, one set of models
which explain the above geometry and the comptonisation origin of
the hard emission in LHS are those based on Advection Dominated
Accretion Flows (ADAF). They are hot radiatively inefficient flows
where most of the energy is advected into the BH (Esin \etal
1998). During LHS, the ADAF takes place between the truncated inner
disc and the BH horizon, and gives rise to a hot optically
thin plasma responsible for the thermal comptonisation of the disc
photons.\\
\indent However, in the recent years, it has become apparent that
in LHS, the BHB become bright in radio and display a clear
correlation between the X and radio luminosities (see Fender \etal
2003). Observations of \x1720 with the ATCA radio telescope have
shown that this source was bright at radio wavelengths during the
secondary outburst (Brocksopp \etal 2004, in preparation), when we
clearly saw the source in the LHS. Simple ADAF models have difficulty
to explain such correlations. Models where a compact jet at the
base of the BH plays a major role in the physical processes of
such systems have been proposed (Markoff \etal 2001). In jet
models of BHB, the high-energy emission seen during the LHS is
interpreted as synchrotron emission from the jets which extends
from radio to
hard X-ray, naturally explaining the correlations observed during the LHS.\\
\indent Besides, the high energy tail observed in HSS or in the
intermediate states is not fully understood. The observed steep
power law (index $\sim$ 2.7) without significance
of a break could also be related to the presence of a non-thermal
component in the accretion flow which has been proposed in the
so-called hybrid thermal/non-thermal models (Zdziarski \etal 2001;
Poutanen \& Coppi 1998). Alternatively, comptonisation on a
population of (thermalised) electrons with bulk motion
(\eg~Titarchuk \etal 1997; Laurent \& Titarchuk 1999) may be
responsible for this component. The fit of \x1720~HSS spectrum we
performed with the BMC model does provide parameters similar to
those obtained in other BHB (Borozdin \etal 1999). However, the
\x1720 spectrum is significant only up to about 80 keV so we
cannot test the predicted presence of a high-energy break expected
at energies
greater than 200 keV.\\
\indent Thanks to the imaging capability and sensitivity of
INTEGRAL, it has been possible to study a faint transient source
in the galactic bulge, to detect a secondary outburst in hard X-rays,
typical of XN, and a spectral transition confirming the probable BH
nature of the object and to obtain  a significant spectrum up to 200 keV.
The detection and study of the other XN of the galactic bulge with
INTEGRAL will possibly provide more data on this kind of objects
and will thus improve our understanding of the physics of BHB.

\section*{Acknowledgements}
MCB thanks J. Paul and P. Ferrando for careful reading and
commenting the manuscript. JR acknowledges financial support from
the French Space Agency (CNES). We thank Brocksopp et al. for
providing a preprint of their paper in preparation. We thank the ESA ISOC and MOC teams for they support
in scheduling and operating the ToO observations of \x1720.
The present work is based on
observations with INTEGRAL, an ESA project with instruments and
science data centre funded by ESA member states (especially the PI
countries: Denmark, France, Germany, Italy, Switzerland, Spain,
Czech Republic and Poland, and with the participation of Russia
and the USA) and with XMM-Newton, an ESA science mission with
instruments and contributions directly funded by ESA member states
and the USA (NASA).

\end{document}